# Analog Computing with Metatronic Circuits

Mario Miscuglio[1], Yaliang Gui[1], Xiaoxuan Ma[1], Shuai Sun[1], Tarek El Ghazawi[1], Tatsuo Itoh[2], Andrea Alù[3], Volker J. Sorger[1,*]

[1] Department of Electrical and Computer Engineering, The George Washington University, Washington, DC 20052, USA

[2] Electrical Engineering Department, University of California, Los Angeles 405 Hilgard Ave., Los Angeles, CA90095-1594, USA

[3] Department of Electrical Engineering, Grove School of Engineering, City College of New York, New York, NY, USA

* Corresponding author sorger@gwu.edu

**Abstract:** Analog photonic solutions offer unique opportunities to address complex computational tasks with unprecedented performance in terms of energy dissipation and speeds, overcoming current limitations of modern computing architectures based on electron flows and digital approaches. The lack of modularization and lumped element reconfigurability in photonics has prevented the transition to an all-optical analog computing platform. Here, we explore a nanophotonic platform based on epsilon-near-zero materials capable of solving in the analog domain partial differential equations (PDE). Wavelength stretching in zero-index media enables highly nonlocal interactions within the board based on the conduction of electric displacement, which can be monitored to extract the solution of a broad class of PDE problems. By exploiting control of deposition technique through process parameters, we demonstrate the possibility of implementing the proposed nano-optic processor using CMOS-compatible indium-tin-oxide, whose optical properties can be tuned by carrier injection to obtain programmability at high speeds and low energy requirements. Our nano-optical analog processor can be integrated at chip-scale, processing arbitrary inputs at the speed of light.

**One Sentence Summary:** A zero-index nanophotonic platform enables post Moore's law analog computing with light, processing data with high throughput and low-energy levels.

**Introduction:**

Current computing tasks are characterized by an elevated level of complexity and therefore require high computational cost. The major criticality of today's digital computing is related to the required computational power, which does not scale well with the problem complexity. Supercomputers are capable of delivering tens of quadrillions of floating-point operations per second, however consuming an amount of energy equivalent to the electrical needs of a city with about 16,000 citizens (*1*). These requirements are growing in time faster than what our society can handle. These intrinsic limitations of digital computing architectures represent an opportunity to engineer entirely different computational paradigms. Developing innovative analog accelerators, which could take the load off from traditional computers by solving specific complex processes, holds the promise to significantly reduce energy consumption and led to the development of next generation heterogenous (i.e. multi-processor) computing systems. Such computing domain specificity enables the possibility of homomorphism where the governing equations are synergistically represented by the underlying hardware. Following this paradigm naturally points to analog processors capable of solving specific and relatively complex problem classes. Analog computers are not a recent invention and are, in fact, well-rooted in human history prior to this digital age, being applied to a vast variety of fields; for instance the ancient Greeks used mechanical tools to predict the astronomical positions, A. Gaudi deployed ropes and mirrors to simulate complex architectural models, and NASA's earlier maned flight missions relied on analog computing systems. While digital processing became dominant over the last 50 years, in recent years the computing landscape has been dramatically perturbed due to the slowing of Moore's law, favoring the advancement of several non-Von Neumann hardware architectures which homomorphically map specific algorithms to hardware. Analog memory (*2–5*), neuromorphic photonics for deep learning applications (*6–10*), optical coprocessors for high speed convolutions (*11–13*), integral equation solvers (*14*) and quantum analog computers (*15, 16*) are some examples of currently explored analog architectures, which can tackle complex tasks more efficiently than a conventional digital processor. The main advantage of analog computers, in particular those which rely on optics and photonics (*17*), is the ability to perform algebraic and integro-differential operations upon continuous signals, enabling efficient non-iterative processing for specific operations. Often analog computations lack a clock (*18*), therefore processes can occur in a completely parallel fashion with no central or distributed memory to access and to wait for.

One of the mathematical tasks that can exploit these paradigms and may greatly benefit from using analog co-processors, and the object of this study, is solving partial differential equations (PDEs). Numerous scientific and engineering problems require the solution of PDEs (*19*), such as problems in thermodynamics (*20*), aircraft design (*21*), and electrical and mechanical engineering (*22*). For solving multidimensional PDEs, current processors require a large number of (iterative) operations, which are computationally-intensive, and, based on the complexity, necessitate considerable amount of memory and power (*23, 24*). Analog computers capable of solving PDEs have been implemented since the 1950s (*19*); primarily using networks or grids of resistive or reactive elements to model the spatial distribution of physical quantities such as voltage, current, and power (in electric distribution networks), electrical potential in space, stress in solid materials, temperature (in heat diffusion problems), pressure, fluid flow rate, and wave amplitude (*25*). However, the complexities of an effective integration of a high-speed programmable and concurrently energy-efficient static-like analog mesh significantly reduced the advancement of this technology.

Nanophotonics is an ideal platform to integrate these efforts, but the short wavelength of light hinders the implementation of scalable, easily reprogrammable lumped technology that characterizes electronic circuits. Here, we introduce a nanoscopic programmable analog processor based on a metatronic nanocircuit board capable of solving PDEs. *Metatronics* was introduced by Engheta a decade ago as the branch of optics that focuses on the control of light at the nanoscale through metamaterial-inspired optical nanocircuitry, and despite finding diverse applications, such as ultra-thin sub-wavelength filters (*26*) for optical signal and performing mathematical operations (*27*), its application as an alternative integrated platform for implementing a compact analog re-programmable processor is still outstanding. The applicability of the metatronic concepts has so far been limited by four main factors: (a) the lack of availability of materials required to realize its individual circuit elements, (b) the absence of controlled processes for straightforward fabrication, (c) lack of re-programmability, 'write', and (d) difficulties in accessing the results, 'read'. Aiming to map a finite difference mesh similar to a network of resistors but in an integrable and ultracompact, nanophotonic platform, we propose a nano-optic circuit based on air groove meshes, engraved in an epsilon-near-zero (ENZ) substrate. We then show that, in absence of considerable losses, the metatronic circuit perfectly matches the solutions obtained with finite difference approaches, analogously to a resistive network. Due to the wide tunability (*28*) of real and imaginary part of permittivity around the ENZ point of indium-tin-oxide (ITO) films, we also explore the possibility of implementing the sub-wavelength circuit using suitably processed ITO. Different deposition conditions are used to tune the ENZ position, and depositing resistive, inductive, and capacitive nanoelements which may lead to top-down monolithically integrated nano-optic circuit boards. The solution accuracy and its scaling, considering the inherent losses of ITO at the ENZ point, are discussed using a finite difference analog method and compared to numerical finite element solutions. We demonstrate and discuss the physical limits of this technology in terms of mesh density and size, and the interplay between undesired losses and wire-coupling. This approach can potentially provide five direct advantages for modern computational needs: first, the metatronic circuit is suitable for direct integration using currently available CMOS compatible materials (can be fabricated using the same manufacturing processes used for silicon electronics), such as monolithic integration of opportunely processed ITO thin layers, allowing the implementation of large-scale metatronic circuit libraries on the same chip. ITO is currently deployed by the high-tech industry at a massive scale, such as in touchscreens or solar-cell electrical contacts. Second, the nano-optic circuit benefits of an ultra-compact footprint comparable to electronics (critical dimension ~10 nm) being much more dense than integrated photonics or RF circuitry. Third, elements of the circuit board could be electrostatically tuned by carrier injection, thus allowing the processor to be re-programmable, enabling to solve a variety of PDEs, including Laplace Equation, Poisson Equation, Diffusion Equation, and the Wave Equation using the same platform. Fourth, the maximum programming latency (~ns) is significantly slower than the transit time (~ps) of the electromagnetic wave in the circuit, thus enabling high speed modulation of the circuit which still would behave as a lumped element circuit where the phase variation in the circuit is still negligible, enforced by the ENZ flattening of the phase and wavelength stretching. Fifth, once the chip is set, the discretized solution is effortlessly computed, and no energy consumption is required for performing operations without any temporization but the transit time of the electromagnetic wave to propagate in the chip. However, there are also potential barriers to overcome in this approach, in particular in terms of the accessibility of the solution, i.e., the direct measurements of the dielectric displacement at each node of the metatronic mesh. In this view, we ultimately discuss the implementation of an all-optical read-out paradigm based on a 'nanophotonic-probe-card that detects the PDE solution in the near-field, providing information of the local dielectric displacement, at given points of the nano-optics circuit, thus allowing to extract the results of the computation in a discretized manner. Our results open a pathway towards the realization of nano-optics accelerators able to efficiently solve complex problem, for those systems such as sensors, where only lesser amount of energy is available, while still demanding high speed.

**Results:**

**Metatronic circuit as a PDE finite difference solver**

Before beginning the description of the nano-optic accelerator, we ought to describe a metatronic mesh and how it compares to a finite difference and lumped resistive meshes. (**Fig. 1A**) We start by considering the steady-state homogeneous Laplace's differential equation, where $f(x,y)$ is the distribution of the physical entity (i.e. PDE solution) of the two-dimensional (2D) domain (i.e. mesh). As a relevant example, we select the heat transport equation because it is widely employed in a variety of mathematical and engineering problems, ranging from machine-learning in graph Laplacian methods (*29*, *30*), to image analysis (*31*, *32*). Then,

$$\nabla^2 f = 0$$

and can be written in the Cartesian 2D space via

$$\frac{\partial^2 f}{\partial x^2} + \frac{\partial^2 f}{\partial y^2} = 0 \quad \textbf{Eq.1}$$

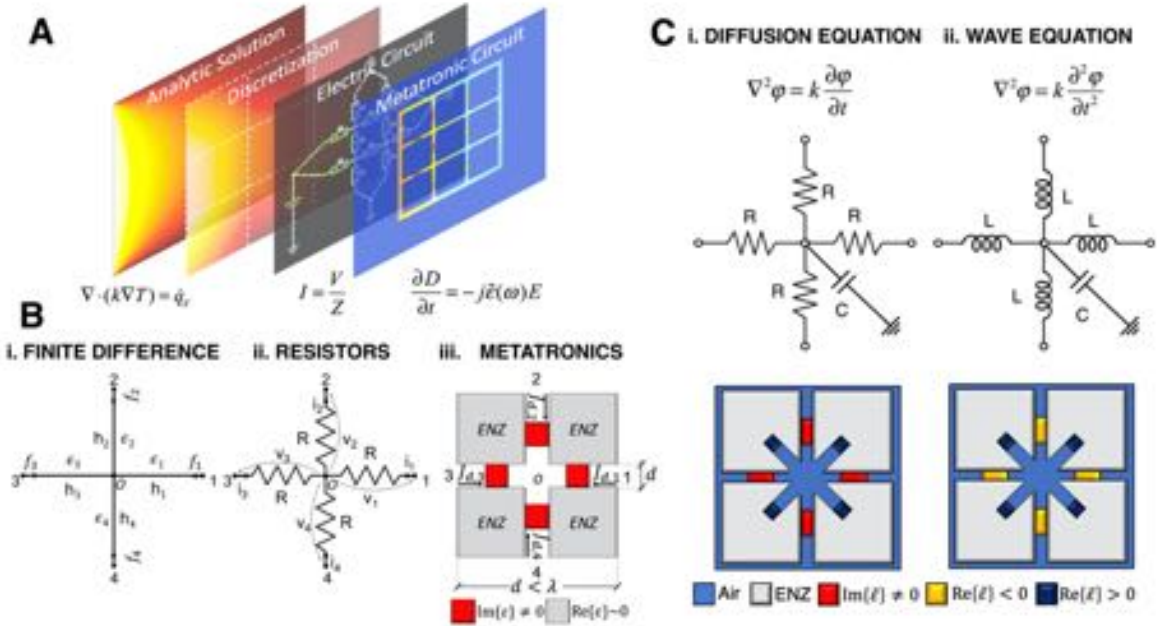

**Figure 1 Solving partial differential equation with metatronics**. (**A**) Sketch of the mapping from a generic partial differential equation to a metatronic circuit, passing through the discretization of finite differences and related Kirchhoff's lumped element equivalent circuit. The schematic shows the solutions of a steady state heat transfer problem with Dirichlet's boundary conditions obtained analytically, and in a discretized fashion with a defined mesh density. The discretized problem is successively converted into a resistive circuit problem which can be ultimately converted into a metatronic circuit in which the nano optical elements are coupled and the light travelling through it is not affected by any phase variation or transit time. (**B**) Node equivalence between finite difference (**i**), resistive mesh (**ii**) and metatronics (**iii**). A finite difference method (i) applied to a heat transfer problem (Laplace equation) is mapped to a purely resistive network and a nano-optic circuit in which the constant point function, $\epsilon_i$, are replaced by resistors in the electrical (ii) and lossy dielectric in the nano-optic circuit (iii), respectively. The finite differences of the function *f*, netting at point *O*, are represented as electrical currents in the electrical circuit (ii) and displacement currents, $J_{D,i}$ (iii) (**C**) Generality of the approach: introducing different nano-elements, either resistive (red block, Im{$\tilde{\varepsilon}$} ≠ 0), capacitive (dark blue, Re{$\tilde{\varepsilon}$}>0) or inductive (yellow, Re{$\tilde{\varepsilon}$}<0), lead to the implementation of diverse time variant second order PDE, such as Diffusion equation (**i**) and Wave equation (**ii**). In this configuration the capacitance is placed in a 'shunt' configuration with respect to the node, split into 4 components for preserving the symmetry of the node.

The same equation may eventually represent a steady state distribution of temperature, as well as stress distributions, potential, and flows. (**Fig 1B.i**) The finite difference method allows current processors to iteratively solve differential equations, such as **Eq. 1**, by approximating them via a "difference equation", in which finite differences approximate the derivatives. Considering equidistant points in the domain (and constant point function $\epsilon_i$), **Eq. 1** can be approximated via the finite difference relaxation method (**Fig. 1B**), at the point $O$ of a mesh, as

$$\nabla^2 \vec{f} \simeq \frac{1}{h^2}\left[\vec{f}(\vec{P_1}) + \vec{f}(\vec{P_2}) + \vec{f}(\vec{P_3}) + \vec{f}(\vec{P_4}) - 4(\vec{f}(\vec{P_0}))\right] \qquad \textbf{Eq.2}$$

The finite difference equation can be straightforwardly compared to the application of Kirchhoff's law to the currents $\vec{f}(\vec{P_i})$ netting at the junction $O$ of a lumped circuit mesh (**Fig. 1B.ii**), where resistances are opportunely scaled ($R = h^2$). Ohm's law, in fact, shows a linear relationship between voltage and current, whilst Kirchhoff's law states that the input and output of current into a node will always be equal (*33*).

According to these equivalences, as previously proposed (*34, 35*), approximate experimental solutions of Laplace's and Poisson's boundary value problems can be obtained using networks of electrical resistances with a rather low tolerance (*34*). The measurement of voltage values at grid points provides the equivalent discrete solution for given resistor values. However, when the parameters of the circuits are programmed at a given rate (i.e., write speed) and its inputs initiated, then an electric mesh might be affected by resistive-capacitive (RC) time delay. Moreover, at high speeds the time at which the signal changes become comparable to the propagation time in the circuit, which therefore makes it into a distributed network. Under these conditions, the circuit would lose the ability to map a finite difference mesh-grid able to solve PDEs. For this reason, a processor capable of solving PDEs using nano-optics circuitry is highly desirable. Here, we exploit the fact that subwavelength nanoparticles (NPs) in the optical domain can be treated as lumped circuit elements, whose impedance is defined in terms of the induced displacement current $J_D$, in response to the local electric field $E$. According to Maxwell's equations, $\tilde{\varepsilon}$, the complex material permittivity, relates these two quantities through

$$J_D = \frac{\partial D}{\partial t} = -j\tilde{\varepsilon}\omega E(\omega),$$

which, for element size considerably smaller than the optical wavelength ($d \ll \lambda$) (*36*), represents an equivalent Ohm's law in the optical domain, enabling the mapping of the resistive circuit. Nevertheless, arranging different NPs with absolute control on permittivity and position is far from easy (*37–39*), and as the size of the circuit grows it is challenging to make sure that the flux of displacement current is funneled to ensure the required circuit connections without phase delays. For these reasons, ENZ substrates are ideal for the implementation of nanocircuit boards, as they enable light to travel through the grooves just like electric currents in copper wiring (*40*) (**Fig. 1B,iii**). Resistors, capacitors, and inductors can be implemented within the air grooves by tailoring the local permittivity values. To map **Eq. 2** in the metatronics paradigm, the resistors are modelled as dissipative dielectrics. Due to the confinement of displacement current in the air grooves, as in an ideal ENZ substrate there is no displacement current leakage in the background, the circuit elements are locally coupled, implying that the Norton/Thevenin equivalents are admissible. In other words, that local variations in the network, generate global effects. Therefore, for a limited functional bandwidth over which the material of the board is ENZ (proportional to the full-width half maximum of the ENZ resonance peak), Kirchhoff's law is satisfied in the mesh, resulting in identical results with respect to a resistive network, reported in **Eq.2**.

A node of a metatronic mesh, equivalent to a resistive one, is represented in **Fig. 1B.iii.** This approach is not limited to resistive networks and consequently to Laplace homogenous equations. Indeed, by using plasmonic nanoparticles (negative permittivity) to construct nanoinductors and dielectric nanoelements (positive permittivity) for nanocapacitors, it is possible to map other time-dependent PDEs applied to a 2D mesh, such as diffusion (**Fig. 1C.i**) or wave-equation (**Fig. 1C.ii**).

In order to prove this concept, we aim to demonstrate that different lumped element functionalities can be achieved in ITO films by simply injecting carriers via electrostatic doping. In detail, an electrical voltage bias can be applied across a capacitor, whose plate is the ITO film, and according to the bias polarity an accumulation/depletion layer is formed at the ITO-oxide interface, increasing/decreasing the ITO's carrier density ($1 \times 10^{19}$-$1 \times 10^{21}$cm$^{-3}$)(*41–43*), which ultimately alters its complex optical properties (increasing/decreasing the imaginary part of the refractive index).

For doing so, we use a straightforward, yet powerful demonstration of a programmable nano-optic circuit based on ITO in which a meta-ring is engraved defining an air loop (**Fig. 2A**). The optical response of the system is obtained by performing a full-wave electromagnetic simulation of an ITO layer. An electric field is driven by a dipole (quantum dot) placed within the channel, forming a directional flow of the displacement current when the ITO layer is in the ENZ range. In this case, we optimized the air trench of width 0.1$\lambda$ to form a closed loop with an average total length of 2πr = 2440nm which is twice the wavelength, and it is sufficiently small for supporting TE$_{10}$-like mode excited by a dipole and concurrently does not experience phase variation within the trenches (**Fig. S1-S2**).

The dispersion of the ITO layer can be electrostatically tuned by varying the carrier concentration (**Fig.2 B**), thus changing the behavior at a given operating frequency. We evaluate the spectral response of the meta-ring, by averaging the field displacement within the air groove for three different carrier concentrations: (i) Depletion (ii) ENZ, (iii) Accumulation (**Fig.2 C,D**).

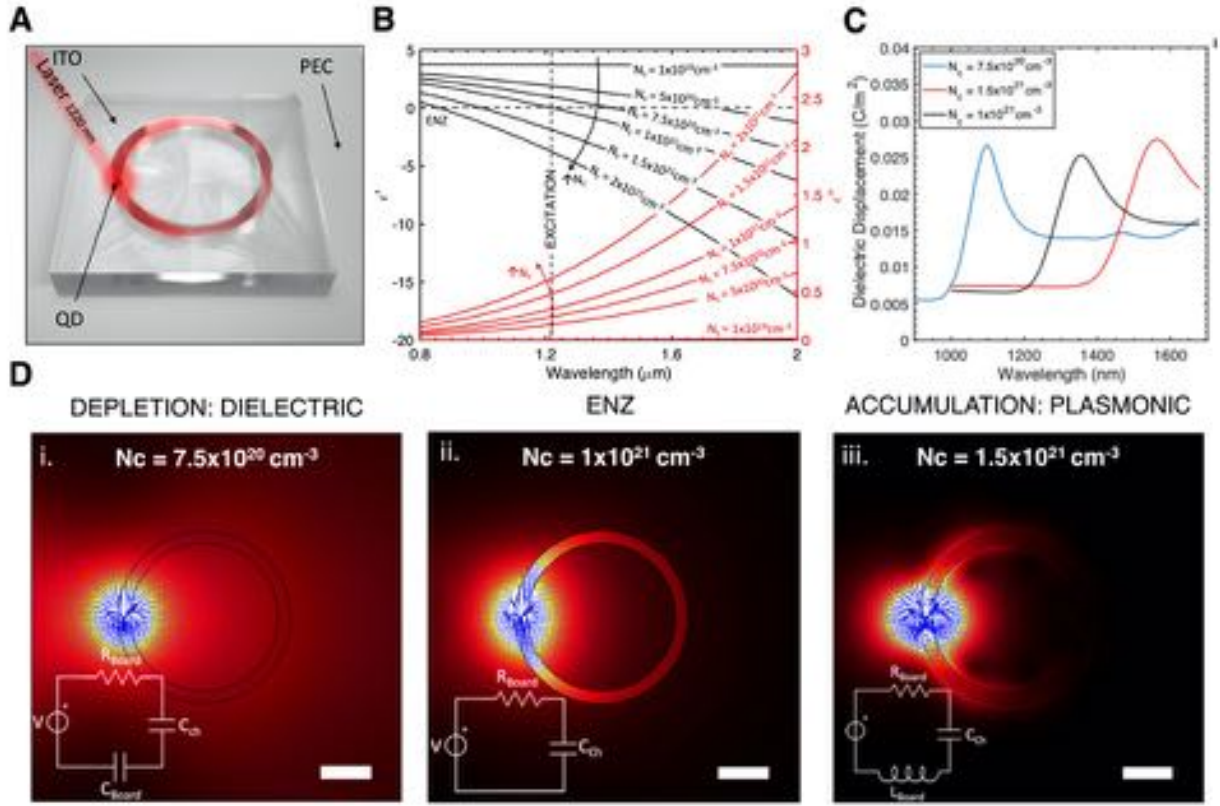

**Figure 2 Epsilon-Near-Zero Tunable Meta-ring**. (**A**) Schematic representation of a meta-ring engraved in an indium Tin Oxide (ITO) layer. The dielectric displacement field within the trench is generated by a dipole, i.e. emitting nanoparticle at 1220nm, which is in resonance with the Epsilon-Near-Zero (ENZ) crossing point of the ITO layer. (**B**) Electrostatically tunable ITO complex permittivity (real part on the left y-axis, imaginary part on the right y-axis) as function of the carrier concentration in both depletion and accumulation regime. The as-deposited film, characterized by spectroscopic ellipsometry, displays a carrier concentration of $1\times10^{21}cm^{-3}$, mobility 38 $cm^2V^{-1}s^{-1}$ and scattering rate $\tau_{sc}$ = 7fs. The model for the "active" dispersion is obtained by fitting experimental ellipsometry data using Drude's model (Details in the Supplementary Materials, see **Fig. S7**). The tuning is envisioned to be obtained by means of carrier injection in a capacitor configuration. The horizontal and vertical dashed lines indicate the ENZ crossing point and the ENZ wavelength of the intrinsic device, respectively. (**C**) Simulated average dielectric displacement within the air groove for different carrier concentrations: depletion ($7.5\times10^{20}cm^{-3}$), intrinsic ($1.0\times10^{21}cm^{-3}$) and accumulation ($1.5\times10^{21}cm^{-3}$). (**D**) Numerically derived displacement field and equivalent electrical circuit for various carrier concentration of the ITO layer. Blue arrows represent the displacement current. (i) for low carrier concentration ITO board behaves as dielectric with low losses; (ii) For a higher concentration ($1\times10^{21}cm^{-3}$) the displacement field is within the air trench; (iii) For furtherly higher carrier concentration, the film is metallic with high losses. Scale bar 200nm. Time domain response of the optical the meta-ring shown in Fig S1 and Movie 1 of the Supplementary Materials.

It is worth observing that for different carrier concentration we obtain a similar response, energy shifted in the same fashion as at the ENZ resonance, indicating a moderately robust response. Additionally, the spectral response is archetypal of a low-pass frequency filter (pass long-wavelength), i.e. integrator-type, since we are considering a voltage drop on a capacitor (dielectric air loop) hosted by a resistive circuit board (lossy ENZ) of the equivalent lumped circuit.

We then fix the excitation wavelength of the film at the ENZ crossing point for the as-deposited case ($1\times10^{21}$ $cm^{-3}$) and take a snapshot of the simulation result of the distribution of the field displacement and current density in the middle plane. In the first scenario (i), when ITO is depleted of its carriers, the material turns dielectric (capacitance, ENZ red-shift), providing relatively low losses ($\varepsilon'(\lambda_{ENZ}) > 0, \varepsilon''(\lambda_{ENZ}) \sim 0$). As a direct consequence, the displacement field in the board hosting the channels becomes non-negligible. (ii) On the other hand, when the excitation matches the ENZ wavelength ($\varepsilon'(\lambda_{ENZ}) = 0, \varepsilon''(\lambda_{ENZ}) > 0$), the air loop behaves as D-dot, in which the displacement current circulates in the loop with relatively low losses. (iii) However, when carriers are accumulated, ITO exhibits a predominantly metallic behavior (inductance, ENZ blue-shifted) characterized by higher ohmic losses ($\varepsilon'(\lambda_{ENZ}) < 0, \varepsilon''(\lambda_{ENZ}) \gg 0$). Equivalent circuit which maps the nano-optic response are displayed as insets (**Fig. 2D i-iii**). For demonstrating the generality of this approach, it is possible to introduce reprogrammable nanoelements for implementing different chip functionality. At ENZ wavelength, the specific response of lumped elements (R, L, C) can be achieved by partially filling the air grooves with ITO nanoelements electrostatically modulated with specific voltage polarity. For illustrative purposes we design the equivalent of an electronic resonator (discussed in Supplementary Materials, **Fig. S3-S4 & Movie 2**) in which energy bounces between a nano-optic inductance and capacitance, thus proving that nano-optic circuits, entirely based on ITO, can be used as programmable oscillators or filters for tuning transmitters and receivers in the telecommunication bandwidth.

**Solution of Laplace homogeneous equation using a metatronic analog processor**

For validating the functionality of the prototyped nano-optic circuit as constituent node of the proposed nanocircuit mesh, we find the solution of a steady-state heat transfer problem in a uniform domain. For demonstration purposes only, we consider the diffusion problem **Eq. 1** with Dirichlet boundary conditions, setting the temperature $T$ at the left edge of a rectangular domain $L \times H$, with $L$ and $H$ width and height of the domain, respectively (**Fig. 3 A,** further details on the derivation provided in the SI). The resistive circuit that solves the finite difference method of a 3x3 mesh is shown in **Fig 3B**, it can be numerically solved using a "spice tool", thus obtaining current and voltage distribution. The boundary conditions that mimic the fixed temperature "potential" at the boundary is a voltage generator $V$, while the heat sink is a simple connection to the ground (relative GND or simply unbiased). The thermal conductivity $k$ of the medium is modelled in the electrical circuit assigning the resistors to be defined as $R = L/kA$. Similarly, we numerically simulate, by using commercially available full-wave simulation software (further details in Methods section of the SI), the electric field displacement and the displacement current in a 3 x 3 metatronic mesh. A snapshot of the simulation result of the distribution of the electric field in the middle plane is depicted in **Fig. 3C**. Here, a strong local electric field, generated by a horizontal dipole, is used to model the heat source, while the ENZ condition is applied to the remainder of the boundaries. In this section, as an illustrative and not limiting example, the permittivity of the circuit board is considered to have negligible losses (*44*) $\varepsilon'' < 0.1 \rightarrow (\tilde{\varepsilon} \simeq 0)$ with an overall size $d$ of 1000 nm, smaller than the operational wavelength $\lambda$ (i.e. $d < \lambda$) to ensure coupling between the nanoelements, as required for conventional electronic circuit concepts at low frequencies (*26, 45*). However, the "spatially static-like" properties of the ENZ substrate, i.e., absence of a significant phase variation in ENZ (**Fig. S1** in the Supplementary Materials, **Movie 1**), essentially relaxes this requirement for the optical nano-circuit board of **Fig. 3C (i)**, for which the total length may become also several free-space wavelengths long (while it is electrically small compared to the very long wavelength in ENZ) (*26*).

Under these conditions, the field lines in **Fig. 3C** (**ii**) highlight that the electric displacement, and consequently the displacement current, fall only within the air grooves, forced by the ENZ conditions in the neighboring area ($\vec{D} \simeq 0$). Solution of the PDE problem from different mesh densities in a circuit with ENZ material with negligible losses $\varepsilon'' \simeq 0$ suggest mesh scalability (**Fig. 3D**). Interestingly, solutions of the PDE adopting a metatronic processor with increasing mesh density keeping the overall circuit dimension maps precisely the solution of a finer mesh in a finite difference approach. This however can be achieved only if the circuit board is characterized by negligible losses, in other words by the absence of a dielectric displacement field in the ENZ circuit board, providing a perfect lumped-element electric-circuit behavior. However, a study of the size and scalability and their impact on the accuracy of the solution of the metatronic processor becomes determinant if the losses in the ENZ circuit board, or the deviation from the ideal zero permittivity condition, are not negligible. Other parameters, such as width of the waveguide (**Fig. S2** in the SI) and smoothness of the bending curves can affect the accuracy of the solution. The undesired influence of these parameters, here not discussed, results in a systematic error that can be compensated or mitigated by accurate and controlled processes.

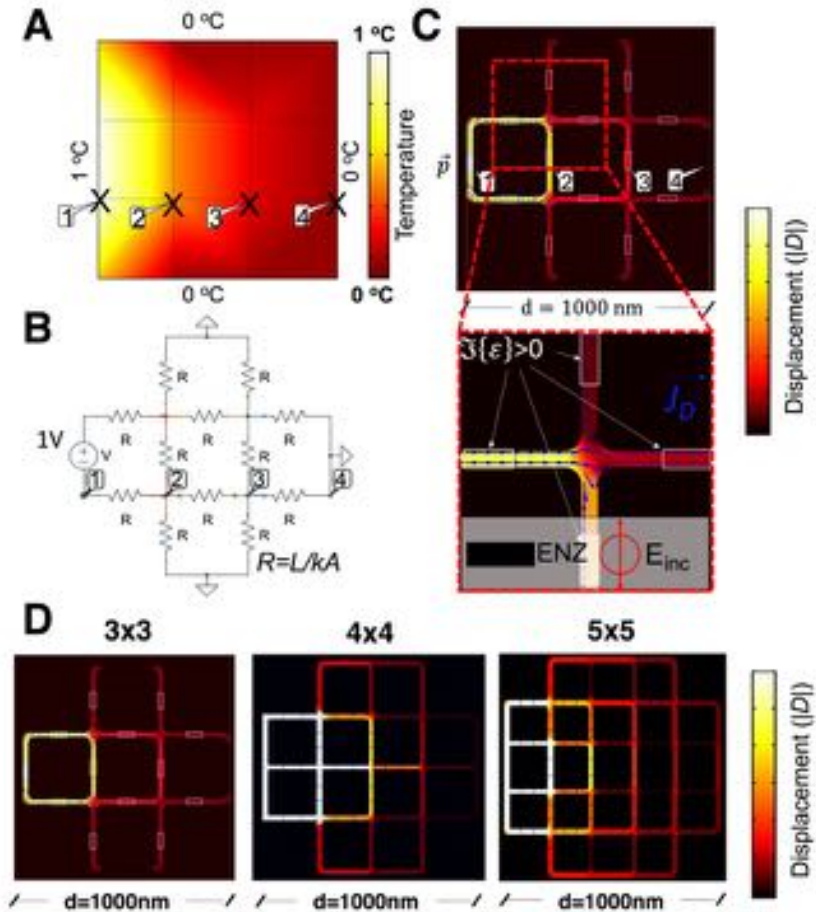

**Figure 3 Heat transfer solution using a metatronic circuit.** Mesh equivalence between finite difference (**A**), resistive mesh (**B**) and metatronics (**C**). Zoom in correspondence of the node, showing the displacement current (blue arrows) following the current path scheme of an electrical mesh. (**D**) Displacement field as a function of the mesh density (3 x 3, 4 x 4, 5 x 5) highlights enhanced accuracy analogously to increased resolution in finite

differences methods. [See Supplementary Materials, **Fig. S5**]. Leakage from the top surface of the structure are limited as shown in Supplementary Materials, **Fig. S6**.

**Monolithic Integration of Reprogrammable Metatronic Processor**

In the past years, a few materials have been considered for fabricating a metatronic circuit board, such as multilayered stacks of thin film (*46*, *47*), NP assemblies and graphene (*48*). However, their large-scale integration is far from easy (i.e., graphene). Here we propose Indium Tin Oxide (ITO) as suitable material for a monolithic integration of the proposed metatronic processor, which enables distinctive functionalities. The advantages of using ITO are manifolds: 1) ITO has a tunable and controllable optical and electrical properties in the NIR according to process parameters (e.g. Oxygen and Argon flow-rate while sputtering, temperature and environment conditions in thermal annealing processes); 2) as previously demonstrated, its optical properties, imaginary and real part of the permittivity, can be electrostatically tuned (*49*, *50*), by carrier injection, thus allowing potentially GHz-fast energy efficient (*49*, *51–53*) re-programmability features on the circuit board [See Supplementary Materials, **Fig. S7-8**]. Moreover, recently, our group achieved a consistent control over ITO optical parameters with respect to the ENZ wavelength as function of sputtering parameter, thus allowing to bridge the technological gap in the implementation of metatronic circuits (*28*). According to our experimental studies (additional information on film characterization provided in the Supplementary Materials), depicted in **Fig. 3B** (**i-iv**), the ITO for the ENZ circuit board could potentially be sputtered with 5 standard cubic centimeters per minute (sccm) Oxygen flow rate, enabling a 200 nm film in ENZ condition at 1550 nm, with non-negligible losses $\tilde{\varepsilon} = 0.3i$ which corresponds to a scattering time $\Gamma = 0.2\text{fs}$. The resistors can be deposited using 20 sccm oxygen flow rate, which yields to $\tilde{\varepsilon} = 1.2 + 0.6i$ and a scattering time of 5fs. The main limitation of the ITO metatronic processor are the unwanted losses in the ENZ circuit board. Because of the losses in the ITO circuit board, the lines of the displacement field are not fully contained in the air grooves, contrarily to the case of an ENZ material with negligible losses, deviating its behavior from a purely lumped-element circuit. This translates into an inaccurate solution which turns to a faster spatial decay compared to the analytical solution. (**Fig.S3** in Supplementary Materials highlights this deviation).

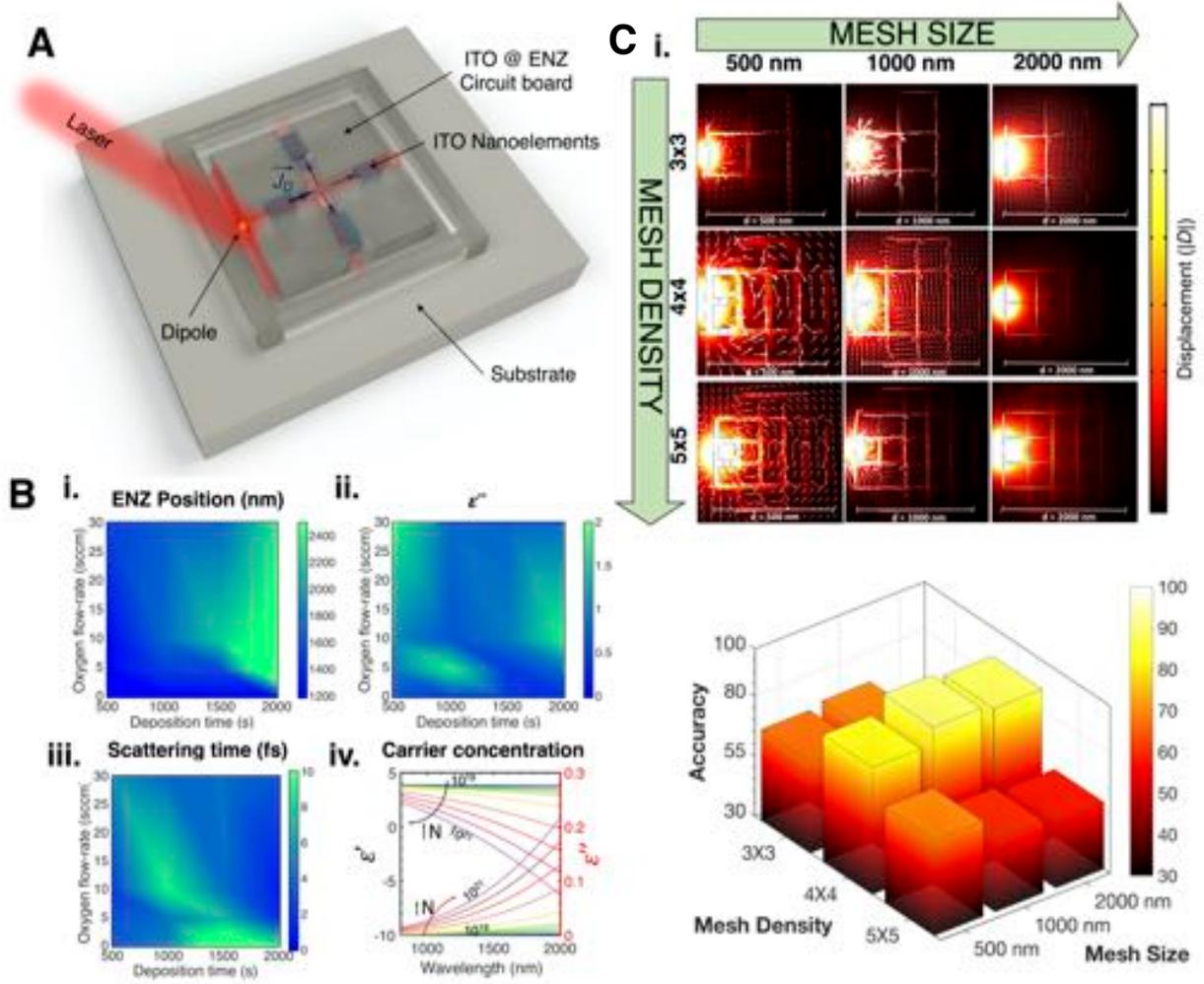

**Figure 4 Monolithic integration of a metatronic circuit board solely based on ITO.** (**A**) Conceptual sketch of a node of a metatronic circuit based on ITO at Epsilon-Near-Zero. A directional antenna is used for coupling the electromagnetic radiation within the trenches, partially filled with dissipative nanoelements which model resistances. A metallic tip of Scanning Near field Optical Microscope is used for collecting the local near field displacement and reconstruct the decay. (**B**) (i-iii) Experimental results obtained through spectroscopic ellipsometry. ENZ wavelength (i), corresponding losses and scattering time $\tau_{sc}$ (ii) as a function of process parameters (oxygen flow-rate and deposition time). (iv) Electrostatic doping. Drude model of ITO, sputtered with an initial electron doping of $10^{19}$ cm$^{-3}$ and $\Gamma = 1/\tau_{sc}$, for an increasing carrier modulation (blue to violet) (**C**) i) Electric field displacement (heat map) in an ITO metatronic processor as function of mesh size (columns) and mesh density (rows). Displacement current is represented as white arrows. ii) Solution accuracy as a function of the mesh density and size, compared to a commercially available solver as a function of mesh density and size.

In presence of non-negligible losses in the ENZ material, the circuit board is not completely insulating, since the displacement current is not negligible:

$$J_D = \frac{\partial D}{\partial t} = \varepsilon'' \omega E(\omega).$$

There are two major coexisting phenomena that impact the accuracy of the solution, both of which depend on the ITO losses (**Fig. 3C (i)**): the first one is a function of the mesh density and the second one of the total physical length of the circuit board. High density (>5 × 5) induces coupling within wires that should not be connected, while the larger physical length (> 2μm) contributes to unwanted dissipation, deviating from the original solution. The accuracy as a function of the number of nodes and physical dimension of the circuit board is given in **Fig. 3C (ii)**. The maximum accuracy (>90%) is obtained for a 1μm size mesh, with a 4 x 4 mesh density. This is achieved thanks to the trade-off between mesh size and density, which minimizes the wire coupling, without extending the wiring length, producing unwanted losses.

As previously discussed, the ITO layer can be in a capacitor configuration, spaced by a thin dielectric, for electrostatic doping, enabling fine tuning of the permittivity values (**Fig. 3B (iv)**) as well as reprogramming the circuit for mapping different problems. The variation of the carrier density via gating in ITO affects both resistance and reactance in the metatronics equivalent circuit, hindering the accuracy of the solution, being imaginary, and the real part of the permittivity. Nevertheless, contrary to the resistive circuit, if either the boundary conditions or the impedances are rapidly "programmed", the nano-optic equivalent circuit is not affected by dispersion, since, unlike electric circuit, here the modulation speed can be arbitrarily large, since signal modulation timescales (>ns) are much larger than the propagation delay of electromagnetic waves across the lumped element. In this case, in fact, the lumped circuit model would hold, since even at 100s of GHz, the timescale at which the carrier signal is modulated does so substantially slower than the time taken by the optical signal to travel through the nano-optics network.

Additionally, besides affecting the accuracy of the solution, the losses affect the power consumption and dissipation. For a processor with a contained size, an approximate solution is always guaranteed. The power consumption from the processor is the summation of the optical power used for exciting the dipole (initiating the processor) and the radio frequency power employed for modulating the carrier density of each lumped elements, i.e., reprogramming the circuit. Concerning the reconfigurability of the processor, recent works showed *few femtojoule efficient* (*45*, *48*) ITO based modulators, potentially operating at high-speed (*53*, *56*) (see SI for GHz-fast experimental ITO modulation). On the other hand, a few milliWatts of optical power are needed for exciting the fluorescent molecule and setting the boundary conditions. Efficient measurement schemes must be used to detect the electric field displacement at each node of the metatronic mesh, avoiding scanning over the sample, e.g., high-resolution tip enhanced near field spectroscopy, to minimize the power used for the detection mechanism.

**Discretized solutions with nano-optic probe card**

In order to sample the electric field displacement signal at the nodes of the metatronic mesh, deep sub-wavelength *near field* microscopy has to be employed with nanometric spatial resolution (*57*, *58*), which can be used for investigating the local near field, breaking the diffraction limit. (**Fig. 5**) However, regular near-field optical microscopes, such as scattering type near-field, are associated with AFM systems, thus requiring long scanning time. For this purpose, we propose a nanophotonic probe card for parallelly reading the values of the local displacement field (**Fig. S7**), similarly to a wafer probe for electrical testing. The reading mechanism is based on an array of dielectric tips characterized by a sub-wavelength aperture at the apex, which collects the local near field radiation similarly to a local near field microscope, allowing for parallel readout.

Aperture SNOM would be preferential with respect to scattering type (s-SNOM), since the former will minimize the coupling between vertical dipoles, i.e., metallic tip, while the latter can introduce second order scattering and a higher degree of uncertainty in the system. Although, a s-SNOM in transmission, which uses a dielectric tip can be employed for measuring the dielectric displacement, in a similar configuration shown in (*59*) with a resolution of a few nm in the lateral dimension (Fig. R4). The s-SNOM would scan the entire ENZ board as well as its trenches in tapping mode. The radiation is conveyed from the bottom using a focused light which by impinging on a QD excites a strong near field within the trenches. The scattered radiation from the tip is collected by a parabolic mirror and provides information of the complex optical properties of the nanoscale region in proximity of the tip. The advantage of using the s-SNOM is that both amplitude and phase of the nearly background-free second tapping harmonic, and certainly background-free in the third harmonic of the local electric near field.

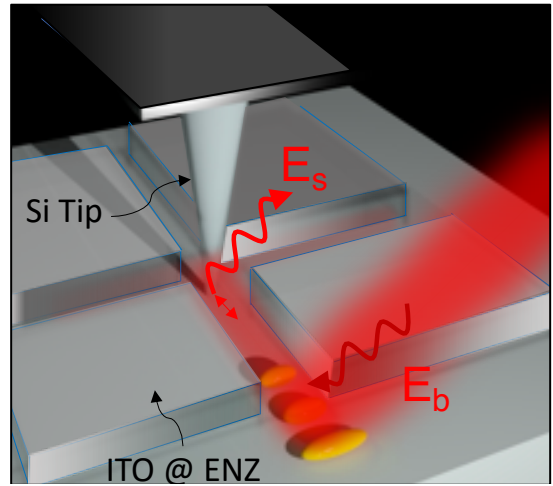

**Figure 5. Schematic representation of a possible experimental demonstration.** Impinging radiation excites a directional antenna, which generates a local near field. The local near field produces a dielectric displacement within a metatronic circuit. The displacement can be probed by the metal tip of a scattering type Scanning Near field Optical Microscope (s-SNOM) or by a Nanophotonic Probe Card (NPC) (Supplementary Materials, **Fig.S9**), which consists of an array of several apertures. (Numerical simulations are presented in the Supplementary Materials, **Fig. S10**)

In conclusion, here we introduced a nano-optic metatronic-based integrated engine, operating in the optical telecommunication band, based on a sub-wavelength Epsilon-Near Zero (ENZ) circuitry. We mathematically prove the equivalence of the mesh current method of a metatronic circuit and finite difference mesh, that solves a steady-state Laplacian equation and by extension other time-variant second order elliptic partial differential equation, by mimicking a lumped-element electrical circuit. Moreover, in contrast with a resistive network, this technology solves partial differential equations with high accuracy (90%), while decoupling circuit mesh upscaling from reprogramming speed. We demonstrate that this metatronic circuit solver can be realized using one single material platform, Indium Tin Oxide, sputtered using controlled process parameters. This analog processor is particularly well-suited to be reprogrammed on both the boundary conditions and network elements, limited in size only by the presence of ohmic losses. Implementing these techniques enables an ultrafast, chip-scale, integrable, and reconfigurable (*60,61*) analog computing processor able to solve partial differential equations at the speed of light.

**Methods**

The numerical simulations of this study were carried out using commercially available software Comsol Multiphysics, we used the frequency dependent solver with a tetrahedron mesh. The dimension of the maximum meshing element of the metatronic board was 1/100 of $\lambda$ and a denser local mesh was considered for the air grooves and the lumped components. The electric field was excited by a point dipole moment of amplitude 1V. The ITO film optical properties (complex permittivity and thickness) and relative sputtering process parameters (oxygen/argon flow rate, deposition time, annealing time and temperature) were taken from ref. (28). The effects of the carrier injection to the dispersion of the ITO film in the near Infrared region was derived using Drude's model. Complex refractive index, scattering rate, ENZ wavelength, initial carrier concentration, thickness and plasma frequency were derived by spectroscopic ellipsometry of the deposited film and used as fitting parameters for the active tuning. Further details regarding the methods used in this study can be found in the Supplementary Materials.

**Acknowledgments:**

**Funding:** This work was supported by National Science Foundation under award number(1748294).

**Author contributions:** V.J.S. and T.E envisioned the idea of a reprogrammable optical computer M.M. and V.J.S. conceived the nano-optic platform for analog computing, V.J.S. and T.E. acquired the funds, and supervised the project. M.M. developed the relevant theories and analyses for the project. M.M. designed the nano-optic circuit and Y.G. and M.M. conducted the spectroscopy ellipsometric experiments. M.M. and X.M. conducted numerical simulations. M.M., A.A., V.J.S. discussed the results and contributed to the understanding, analysis, and interpretation of the results. M.M. wrote the first draft of the manuscript, and V.J.S., A.A. T.I., and T.E. contributed to writing subsequent drafts of the manuscript.

# Supporting Online Materials for

## Analog Computing with Metatronic Circuits

Mario Miscuglio[1], Yaliang Gui[1], Xiaoxuan Ma[1], Shuai Sun[1], Tarek El Ghazawi[1], Tatsuo Itoh[3], Andrea Alù[2], Volker J. Sorger*

**Affiliations:**

[1] Department of Electrical and Computer Engineering, The George Washington University, Washington, DC 20052, USA

[2] Department of Electrical Engineering, Grove School of Engineering, City College of New York, New York, NY, USA

[3] Electrical Engineering Department, University of California, Los Angeles 405 Hilgard Ave., Los Angeles, CA90095-1594, USA

* Corresponding author sorger@gwu.edu

This PDF file includes:
- Meta-ring Temporal Analysis
- Meta-Resonator
- Details on ITO Film Characterization
- FDM and Metatronic Circuit: comparison between discretized solutions.
- Mode leakage outside the trenches.
- Launching propagating mode within the air-trenches of the ENZ board
- Solution of a generic elliptic equation using FDM and mapping onto a metatronic engine

- **Figure S1.** Epsilon-Near-Zero Tunable Meta-ring temporal response.
- **Figure S2.** Impact of the width of the trenches in Epsilon-Near-Zero Tunable Meta-ring.
- **Figure S3.** Meta-resonator.
- **Figure S4.** Magnetic field of the meta-resonator.
- **Figure S5.** Mode leakage outside the air trench in the ENZ board.
- **Figure S6.** Comparison of the discretized results.
- **Figure S7.** ITO dispersion and doping.
- **Figure S8.** Experimental characterization of a sub-wavelength ITO Plasmonic Mach-Zehnder Modulator on Silicon Photonics.
- **Figure S9.** Nanophotonic probe card.
- **Figure S10-13.** Map of an elliptic PDE on the metatronic engine
- **Movie 1:** Harmonic extension of the Dielectric displacement in a ITO-based Meta-ring vs carrier density.
- **Movie 2:** Harmonic extension of the Dielectric displacement in an RLC meta-resonator.

# Meta-ring Temporal Analysis

In reference to Fig.2 of the main manuscript, here we highlight the temporal response of the ITO meta-ring to a resonant dipole with the dipole moment positioning being parallel to the waveguide walls, placed within the air groves of the board for different carrier concentrations of the film. (**Fig. S1 A-C**). In case of a depleted ITO film ($N_c=7.5\times10^{20}$ cm$^{-3}$), due to a shift towards lower energy of the ENZ position, the board behaves as a dielectric material, not allowing confinement of the light in the air trenches of this ENZ circuit board, (**Fig. S1 A**). For higher carrier concentration ($N_c=1\times10^{21}$ cm$^{-3}$), at ENZ, the light is forced to be funneled within the trenches engraved in the ENZ board. The dielectric displacement highlights the absence of a significant phase variation of the signal, reinforcing the spatially static-like characteristic of the ENZ board despite the circuit being 2 times larger than the wavelength ($2\pi r$= 2440 nm, $r$ = ring radius). When the ITO layer is heavily doped (hdITO) ($N_c=1.5\times10^{21}$ cm$^{-3}$), the board becomes metallic ($\varepsilon' < 0$), and the nanostructure becomes an in-plane (xy) Metal-Insulator-Metal (MIM) lossy structure, constituted by a hdITO-air-hdITO layers (**Fig. S1 C**). The dielectric displacement exhibits the classic behavior of an electromagnetic wave associated to a surface plasmon polariton, which propagates at the metal-dielectric interfaces in counter-phase, as can be seen by the propagating surface-edge wave. **Movie 1** shows the harmonic extension of the meta-ring, as response to a dipole excitation.

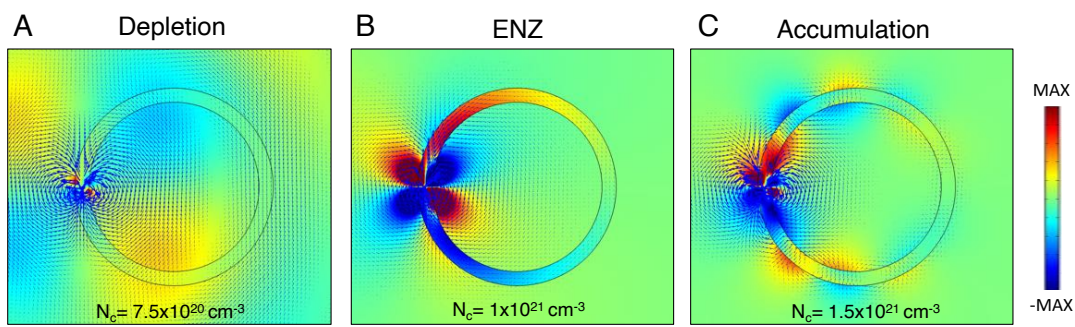

**Figure S1. Epsilon-Near-Zero Tunable Meta-ring temporal response**. Numerically derived x-component of the displacement field induced by dipole placed within the air trenches. Blue arrows represent the displacement current. (A) for low carrier concentration ITO board behaves as dielectric with low losses; (B) For a higher concentration ($1\times10^{21}$cm$^{-3}$), at ENZ, the displacement field is within the air trench; (C) For furtherly higher carrier concentration, the film is metallic with high losses. Time domain response of the optical the meta-ring can be appreciated in **Movie 1** of the Supporting Online Material.

In this case, we optimized the air trenches of width $0.1\lambda$ to form a closed loop with an average total length of $2\pi r$ = 2440nm which is twice the wavelength, and it is sufficiently small for supporting TE$_{10}$-like mode excited by a dipole and concurrently does not experience phase variation within the trenches (**Fig. S2**).

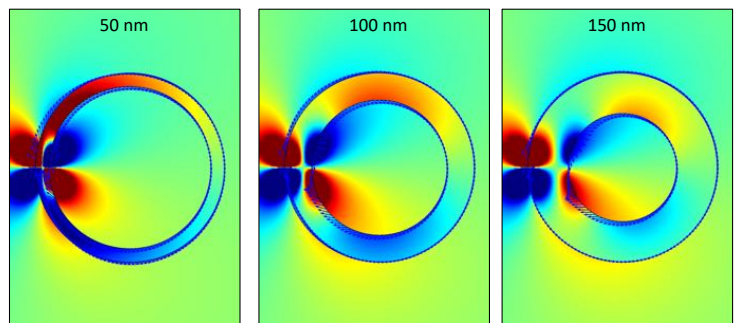

**Figure S2. Impact of the width of the trenches in Epsilon-Near-Zero Tunable Meta-ring.** Dielectric displacement generated by an electric dipole placed within air trenches of different size (50 nm, 100 nm, 150 nm). For larger trenches higher edge-modes emerge.

## Meta-resonator

Following the characterization of the meta-ring, next we introduce nano-optic elements within the air-trench, thus adding resonator tunability to the system towards adaptation functionality of the nanoscale 'filter' (**Fig. S2 A**). The nanoelements are physically realized with the same basic material as the ENZ circuit board, namely ITO, however, bare a different carrier density to exhibit an *RL* or *RC* element. Note, this is realized either statically such as via changing the material process conditions during the ITO sputter recipe, and/or actively (i.e. electro-static tuning) towards deliver element-wise independent tunability of sub-wavelength small (i.e. lumped) elements (**Fig. S2 B**). As a demonstration, here we consider one element to be injected with carriers (*L*) and the other one depleted (*C*), resulting into an *LCR*-like behavior at the ENZ wavelength. Note, since ITO's index tunability originates from free carrier modulation and the carrier densities are rather high (even in the depleted case, i.e. depleted here refers to a reduced carrier concentration, not void-of-carriers), an EM wave impeding ITO will always experience a non-zero imaginary optical index, this adding loss, which electrically resembles a resistance *R*. Thus, a pure *C*, or pure *L* element is not simply possible in ITO unless gain or loss-compensation methods are used. Our model nanocircuit considered here acts as a nano-optic resonator storing energy oscillating at the circuit resonant frequency calibrated to be at ENZ (**Movie 2**).

In an electrical *RLC* circuit, the voltage drop on a capacitance and inductance are represented by the following rational functions:

$$V_c(\omega) = \frac{1}{-\omega^2 LC+1} V_{in}, \quad V_L(\omega) = \frac{\omega^2 LC}{\omega^2 LC-1} V_{in}.$$

The function $V_c(\omega)$ is characterized by two poles in $\omega = 1/\sqrt{LC}$, while $V_L(\omega)$ has two zeros at 0 frequency and 2 poles in $\omega = \frac{1}{\sqrt{LC}}$.

Besides the correct mapping of the optical circuit at ENZ, the anticipated Low Pass (LP)/High Pass (HP) frequency response of the electrical circuit is not perfectly mapped by the dielectric displacement in the ITO nano-optics circuit for each wavelength (**Fig. S2 D**). This is primarily due to 3 factors:

**(i)** ITO nanoelements are affected by a strong dispersion are functions of the wavelength unlike the electrical counterpart. More in detail, the optical impedances can be written as:

$$Z_c(\omega) = (-i\omega C(\omega))^{-1} \quad with \quad C = \varepsilon'_c(\omega)\varepsilon_0 t/l \quad being \quad \tilde{\varepsilon}_c(\omega) \sim \varepsilon_c'(\omega)$$

$$Z_L(\omega) = i\omega L(\omega) \quad with \quad L = -l/(\omega^2 \tilde{\varepsilon}_L(\omega)\varepsilon_0 t)$$

considering *l* and *t*, length and thickness of the nano-optic element, respectively and $\omega$ the optical angular frequency. It can be noticed that the dielectric displacement (and consequently the electric field) within the capacitance is in counter phase with respect to the phase in the inductance (**Fig. S2 C**), as expected.

**(ii)** The static-like behavior of the metatronic board is guaranteed only at ENZ

**(iii)** Optical losses in the circuit boards.

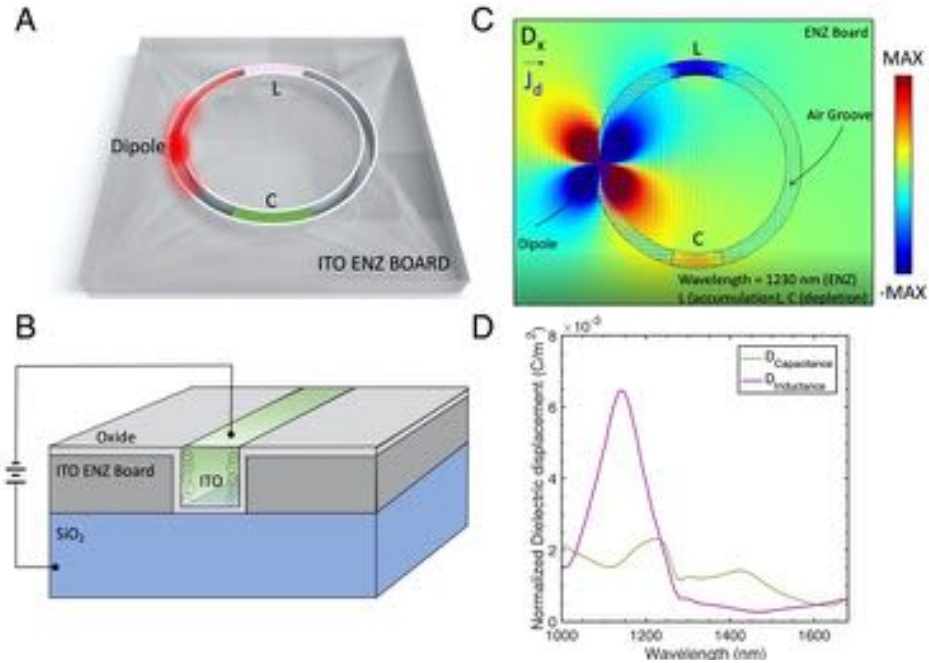

**Figure S3. Meta-resonator**. Schematic representation of the nano-optic circuit mimicking an *RLC* optical lumped circuit at ENZ (**A**) and related bias mechanism for modulating the carrier density, thus reprogramming the optical circuit to exhibit a particular frequency response and acting as a programable "'filter" (**B**). The functionality of each nanoelement is obtained through electrostatic doping in a capacitor configuration. Carrier depletion ($7.5 \times 10^{20}$ cm$^{-3}$) prompts a capacitor behavior, while carrier accumulation ($1.25 \times 10^{21}$ cm$^{-3}$) induces an inductance behavior. (**C**) Simulation results for the x-component of the displacement field induced by dipole placed within the air trenches. Blue arrows represent the displacement current. Normalized dielectric displacement spectral response at the capacitance and inductance.

The simulation results show that the z-component of the magnetic field, Hz (**Fig. S4**) in the middle plane of the circuit is negligible within the trenches (like in wires of an electrical circuits) while almost being constant within the loop. The small discrepancy to what was reported in Ref. (1) is ascribed to the optical losses in the ENZ circuit board.

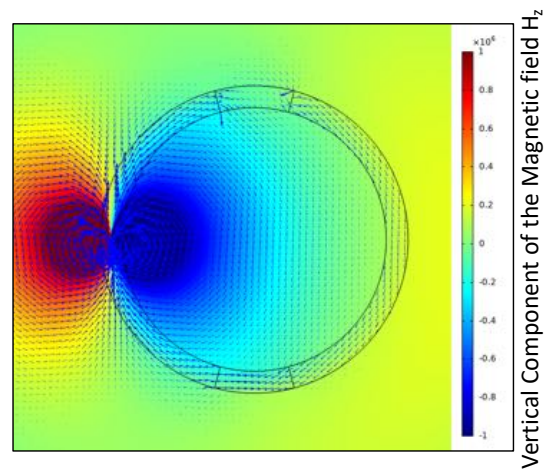

**Figure S4. Magnetic field of the meta-resonator.** Simulation results for magnitude of z-component of magnetic field on the middle plane of the nano-LC tuner.

# FDM and Metatronic Circuit: comparison between discretized solutions.

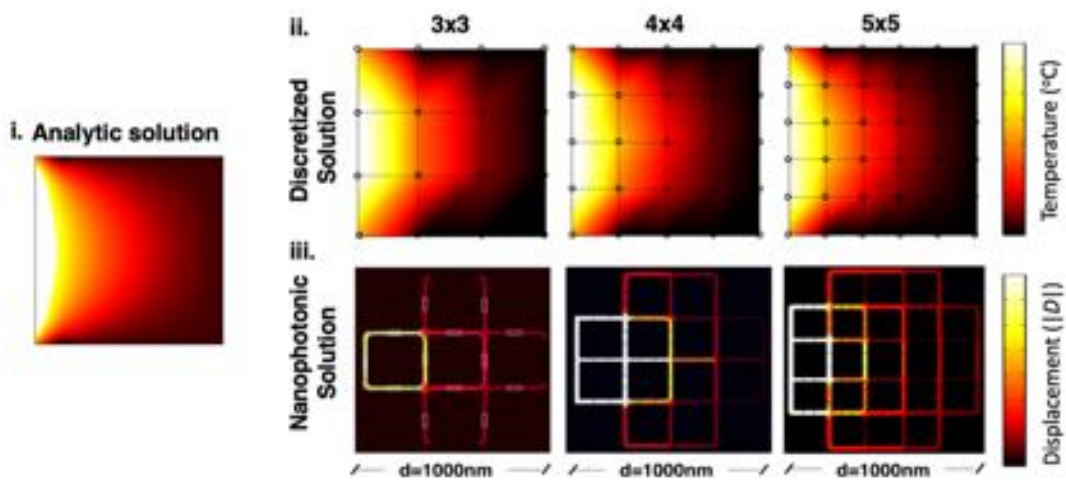

**Figure S5. Comparison of the discretized results.** Comparison between the analytic solution of the Laplace equation, the discretized solution obtained through numerical simulations (FDM) and the normalized dielectric displacement at each node for different mesh densities (3x3, 4x4 and 5x5).

# Mode leakage outside the trenches.
As shown by Li et al in (*5*), it is preferable to cover a nano-optic waveguide with a Perfect Electric Conductor, which would avoid the leakage in free space. Nevertheless, when the mode is generated within the air trenches, exploiting near field of an electric dipole (Quantum Dot), the radiation is conveyed without radiating out excessively (**Fig S6**). We include 3d full-wave FDTD numerical simulations (Lumerical) which showcase the low propagation in the air (<1 dB) compared to the propagating field within the trench.

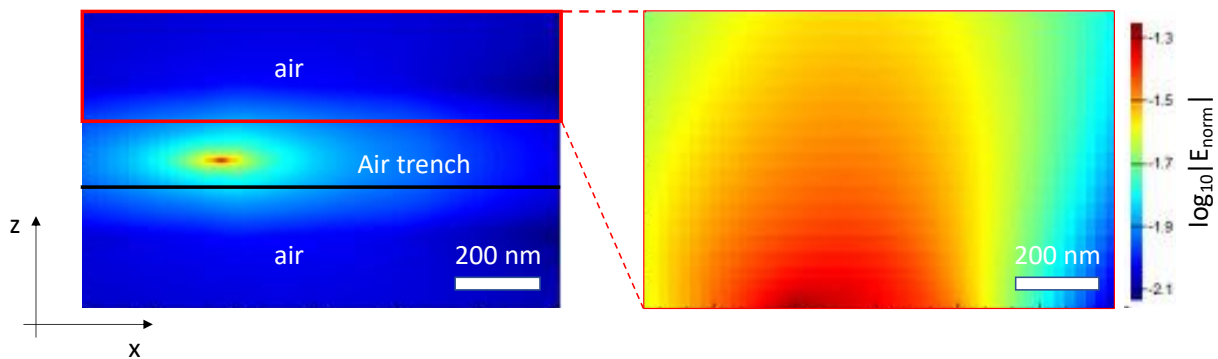

**Figure S6. Mode leakage outside the air trench in the ENZ board.** Cross section view of the normalized electric field map of the propagating mode within the slot waveguide and related cross-sectional view of the ENZ card of the electromagnetic radiation of the air trench

# Details on ITO film characterization and speed measurements.
The properties of the ITO films used in the numerical simulation, are extracted from measurements performed on experimentally deposited films in Ref. (*1*). Ultra-thin films of ITO were deposited on a cleaned Si with a nominal 300 nm $SiO_2$ on it (1cm x 1cm) at 313K by reactive RF sputtering using Denton Vacuum Discovery 550 Sputtering System. The target consisted of 10% $SnO_2$ and 90% $In_2O_3$ by weight. The ITO films were prepared with the same Argon flow-rate which is 40 sccm and different oxygen flow-rates (0 sccm, 5 sccm, 10 sccm, 20 sccm, and 30 sccm). The influence of the deposition time has also been studied. The vacuum setpoint was 5 Torr before deposition and the target was pre-sputtered with the same deposition condition. RF bias voltage was 300 V and RF bias was 25 V. After all parameters reached their setpoints, deposition began. After deposition, samples were annealed in a sealed chamber filled with $H_2$ and $N_2$ at 350º C for 15 min in order to activate the carriers.

The ITO film properties have been obtained through spectroscopic ellipsometry measurement using J.A. Woollam M2000 DI, which covered wavelength from 200 nm to 1680 nm. Analysis of the data was used CompleteEASE to extract thickness, complex optical constants, and other electrical parameters. Our approach consisted in first fitting the transparent region of the ITO film using Cauchy model thus obtaining a reliable estimation of the thickness. Secondly, we fitted the data using B-spline model and subsequently we expanded the fitting region from transparent region to the entire wavelength region. Successively we re-parameterized the data using different oscillators in the GenOsc, i.e. Drude, Cauchy-Lorentz and Lorentz oscillators.

For the values of the actively tuned permittivity, used in the electromagnetic simulations of the meta-ring and *RLC* resonator, we use ellipsometry data of the film for finding to a Drude model, thus obtaining an estimation of initial, as deposited, carrier concentration, scattering rate and Fermi energy of the film. Afterwards these parameters are used in the following formula for obtaining the optical response for electrostatically tuned films Ref. (*2*):

$$\varepsilon(\omega, \text{ITO}) = \varepsilon_\infty - \frac{N_c e^2}{\varepsilon_0[m^*(ITO)\omega^2 + ie\omega/\mu(ITO)]}$$

where $N_c$ is the carrier density, $e$ the electron charge, $\varepsilon_0$ the vacuum permittivity, $m^*(ITO)$ and $\mu(ITO)$ are the electron effective mass and mobility, which vary with electron temperature. The plasma frequency $\omega_p = N_c e^2/[\varepsilon_0 m^*(ITO)]$ decreases with temperature and the scattering rate $\gamma = e/[m^*(ITO)\mu(ITO)]$ increases.

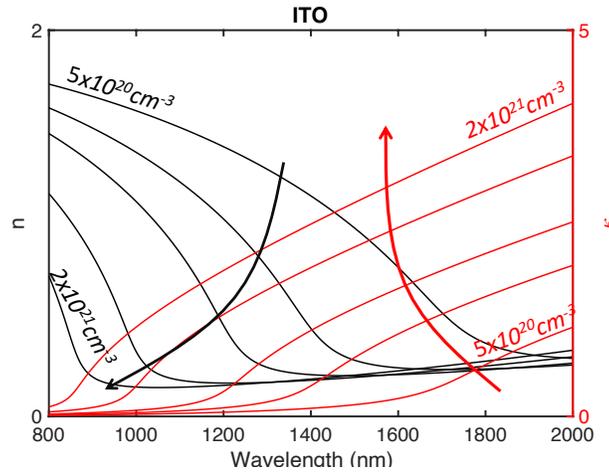

**Figure S7. ITO dispersion and doping.** Complex refractive index (real and imaginary parts, n and $\kappa$) of the ITO film vs wavelength for different carrier concentration.

To corroborate the statements in the main manuscript regarding reconfiguration speed and energy efficiency, we report on a recently prototyped ITO Plasmonic Mach-Zehnder Modulator on Silicon Photonics with a footprint of just 1.4μm, which can operate at a speed of 1.1 GHz (**Fig. S6 A**) with contained losses and characterized by as low as 11 pJ/bit energy consumption Ref. (*3*). Deployment of high-k gate dielectrics including thickness of the oxide scaling can enable the realization of metatronic circuits which only require a few fJ/bit to be reprogrammed at GHz speed.

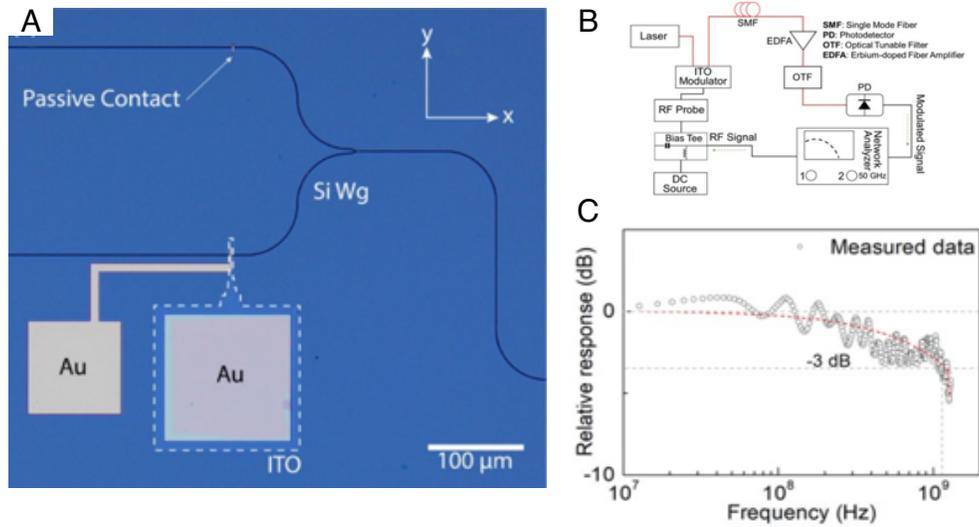

**Figure S8. Experimental characterization of a sub-wavelength ITO Plasmonic Mach- Zehnder Modulator on Silicon Photonics.** (**A**) Optical microscope image of the sub-λ ($L_d$ = 1.4 μm) modulator. (**B**) Experimental high-speed set-up. Frequency response ($S_{21}$) is obtained by generating a low power modulating signal (0 dBm) with a 50 GHz network analyzer; a bias-tee combines DC voltage bias (6 V) with the RF signal. RF output from the modulator is amplified using broadband EDFA (~35dB), an optical tunable filter enhances the signal integrity and reduces undesired noise by 20dB. The modulated light is collected by a photodetector with a single mode fiber.; and (**C**) measured small-signal response, $S_{21}$ of the modulator establishing a -3 dB bandwidth of 1.1 GHz.

## Launching propagating mode within the air-trenches of the ENZ board

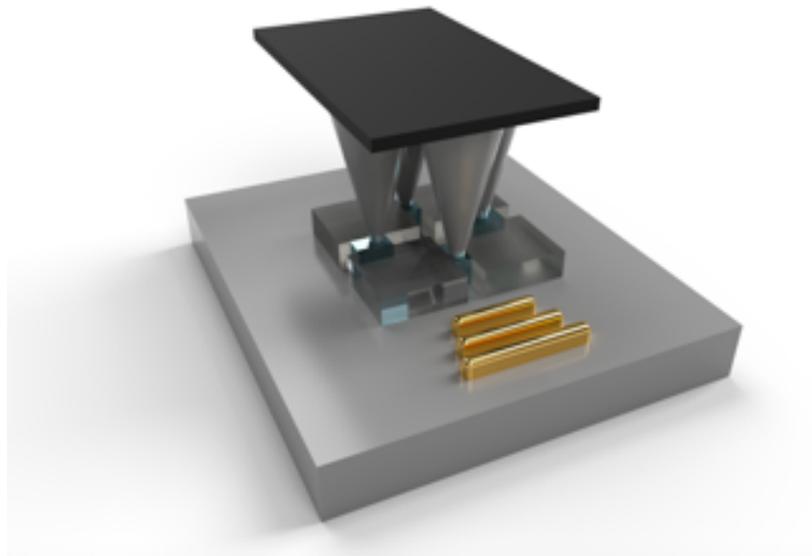

**Figure S9. Nanophotonic probe card.** The displacement can be probed by a Nanophotonic Probe Card (NPC) (Supporting Online Material Fig.S7), which consists of an array of several apertures, which can read in parallel the dielectric displacement at each node.

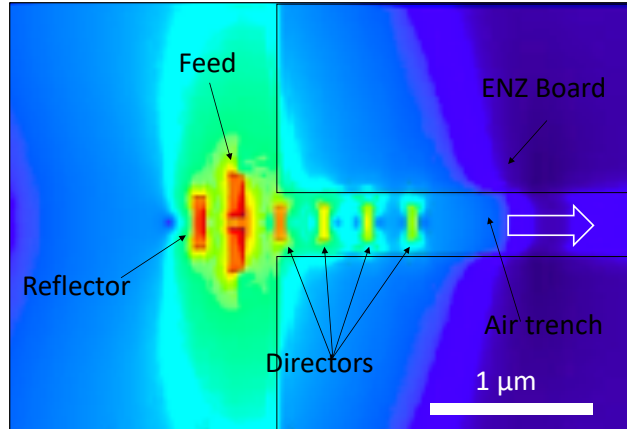

**Figure S10. Nanoantenna for launching electric near field within the air trenches.** Normalized Electric field associated to light launched by a Yagi-Uda antenna. The antenna is illuminated with a focused beam at the ENZ wavelength of the board which is also the resonant frequency of the antenna. A reflector avoids back reflection and the directors of the antenna, which are placed within the air trench of the ENZ board) are responsible for a highly directional propagation within the air trenches in ENZ board. Radiation is also lost in free space and in the substrate.

Having an antenna or a grating coupler at the edge of the ENZ board will not be an effective way to launch a propagating mode within the air trenches of the ENZ board, primarily due to reflection because of the large impedance mismatch. We used a Yagi-uda antenna, consisting of a feed (resonant at the ENZ wavelength of the board), a reflector for avoiding back propagation, and a series (x4) of that directors with the air trench for launching the propagating mode. The antenna could electrically launch an optical signal using tunnel junction.(*4*)

## Solution of a generic elliptic equation using FDM and mapping onto a metatronic engine

Considering non homogenous, second order, elliptic Partial Differential Equation:

$$2\frac{\partial^2 f}{\partial x^2} + 7\frac{\partial^2 f}{\partial y^2} = 5 \qquad \text{Eq.1}$$

The second order spatial derivatives can be written in terms of incremental ratios (central differences) at the mesh node of the discretized space (discretization step in x and y $\Delta x$ and $\Delta y$):

$$\frac{\partial^2 f}{\partial x^2} = \frac{f_{i+1,j} - 2f_{i,j} + f_{i-1,j}}{\Delta x^2}$$
$$\frac{\partial^2 f}{\partial x^2} = \frac{f_{i,j+1} - 2f_{i,j} + f_{i,j-1}}{\Delta y^2} \qquad \text{Eq.2}$$

Assuming $\Delta x = \sqrt{\frac{2}{7}}\Delta y$, Eq.1 can be written as the Laplacian difference equation:

$$2\frac{f_{i+1,j} - 2f_{i,j} + f_{i-1,j}}{\Delta y^2 \left(\sqrt{\frac{2}{7}}\right)^2} + 7\frac{f_{i+1,j} - 2f_{i,j} + f_{i-1,j}}{\Delta y^2} = 5 \qquad \text{Eq.3}$$

Which leads to the simplified difference equation:

$$\frac{(f_{i+1,j} - f_{i,j}) + (f_{i-1,j} - f_{i,j})}{\Delta y^2} + \frac{(f_{i+1,j} - f_{i,j}) + (f_{i-1,j} - f_{i,j})}{\Delta y^2} = \frac{5}{7} \qquad \text{Eq.4}$$

Applying the same approach to all the N nodes of the mesh we obtain a system of equations, characterized by a sparse matrix, that can be solved iteratively with Liebman or Gauss-Seidel method, assuming initial condition at each node being $f_{i,j} = 0$. At node 1,1, for instance, $f_{1,0}$ and $f_{0,1}$ are boundary conditions (**Fig. S9**).

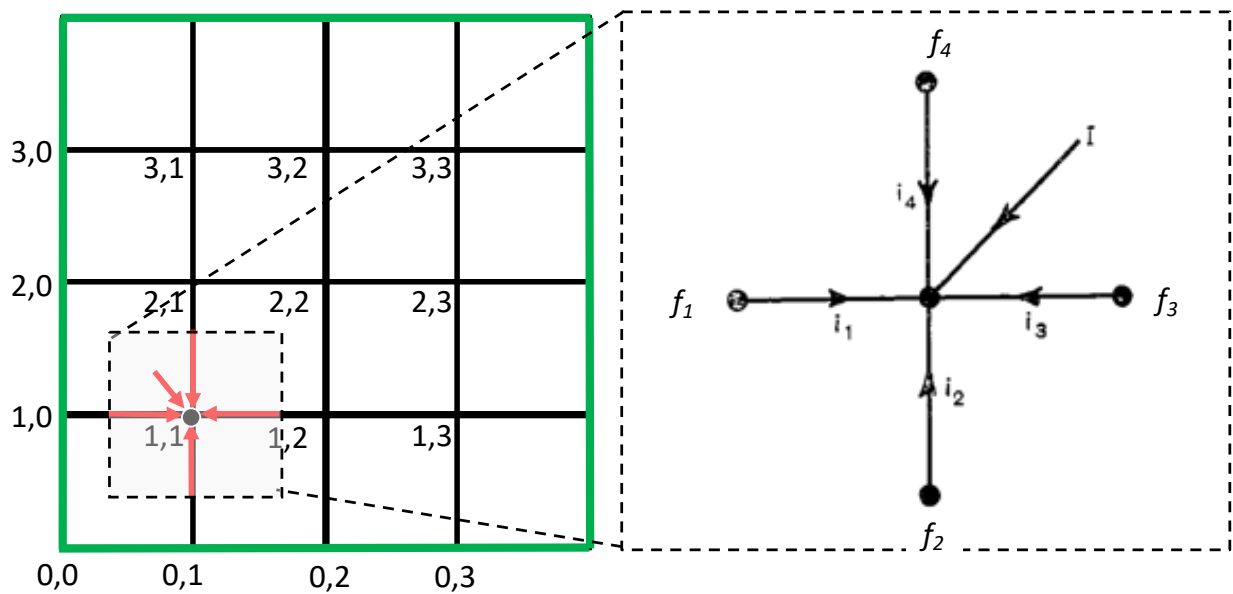

*Figure S11. Discretization mesh and related FDM node for a second order PDE.*

The difference equation (Eq. 4) is equivalent to the Kirchhoff's equation at the node of the electrical circuit of equation:

$$\frac{(V_1 - V_0)}{R_1} + \frac{(V_2 - V_0)}{R_2} + \frac{(V_3 - V_0)}{R_3} + \frac{(V_4 - V_0)}{R_4} = -I \qquad \text{Eq.5}$$

Where $R_1$, $R_2$, $R_3$ and $R_4$ are equal to $\frac{1}{\Delta y^2} R_0$. Considering the discretization step size $\Delta x = \sqrt{\frac{2}{7}} \Delta y \sim 0.53 \Delta y$. With external source at each node equal to $\frac{5}{7 R_0}$.

In this particular case, the nano-optic circuit will be characterized by an anisotropic step size of the mesh ($\Delta x \sim 0.53 \Delta y$) and nano-elements within air trenches characterized by the same impedance (**Fig. S9**), or longer nanoelements in the x-direction. Additionally, an external source (electric quadrupole) needs to be placed at each point of the mesh. In the case of an ENZ circuit board characterized by zero losses, the dielectric displacement represents the discretized solutions using FDM. In case of ITO, considering uniform losses due to ENZ board, as demonstrated in the main manuscript it is possible to obtain an accurate solution.

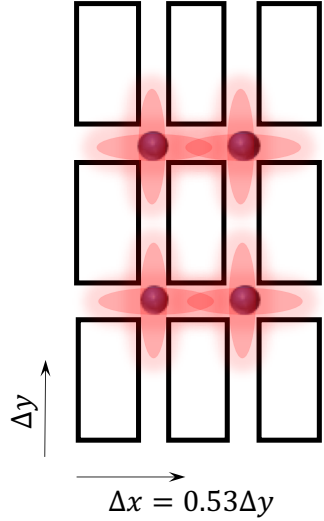

*Figure S12. Passive implementation of the nanooptics circuit for an arbitrary PDE.* *Passive implementation of the nano-optic circuits which avails itself of an anisotropic mesh for solving the assigned PDE.*

Alternatively, without using the anisotropy of the mesh density, and exploiting the programmability of the nano-optic component, it is possible to realize an accurate solution of the PDE. To do so by forcing the optical losses in the y direction to be roughly 2 times larger than the optical losses in the x direction for each node. Considering that the values of the resistances can be normalized, for the nano-optic element within the trenches we can use ITO with initial very low losses (not annealed ITO, with an initial carrier concentration of $1\times10^{19}$ cm$^{-3}$ and $\tilde{\varepsilon}$ =3.74+0.003j) and electrostatically doped only the elements in the y direction ($5\times10^{19}$ cm$^{-3}$ and $\tilde{\varepsilon}$ =3.6+0.012j losses). In this case starting with initial low losses the nanoelement will undergo a significant variation of $\varepsilon''$ without a concurrent variation of the $\varepsilon'$. Although in this case each node will be characterized by similar capacitances in all the directions (**Fig. S11**).

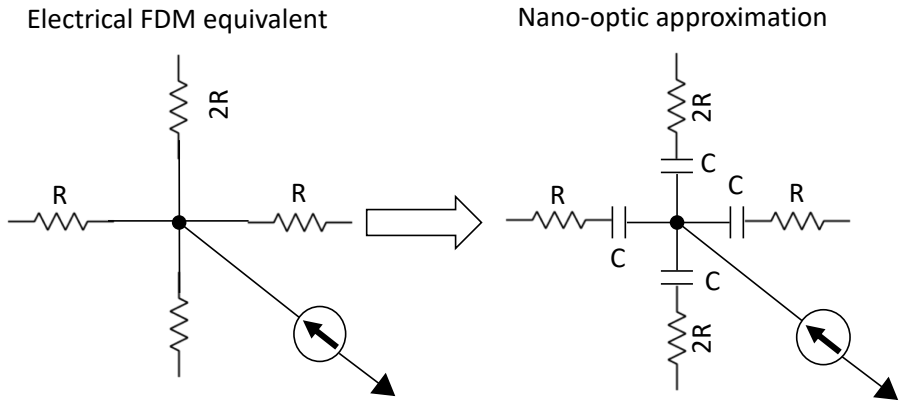

*Figure S13. Equivalent circuits of the elliptic PDE.* *Active implementation of the nano-optic circuits which avails of electrostatic modulation of the permittivity.*

Now, assuming the dimension of the nanoelement being comparatively smaller than the effective wavelength in the trenches, equation 5 becomes:

$$\frac{(V_1 - V_0)}{Z_1} + \frac{(V_2 - V_0)}{Z_2} + \frac{(V_3 - V_0)}{Z_1} + \frac{(V_4 - V_0)}{Z_2} \sim -I \qquad \text{Eq.6}$$

Where $Z_1 = R + \frac{1}{j\omega C}$ and $Z_2 = 2R + \frac{1}{j\omega C}$ are the impedances of the circuit in the x and y direction, respectively. In terms of nano-optic elements this is equal to $Z_1 = -i\omega \frac{t}{l}\varepsilon_0 \varepsilon_{\text{eff},1}$ and $Z_2 = -i\omega \frac{t}{l}\varepsilon_0 \varepsilon_{\text{eff},2}$, where $\varepsilon_{\text{eff},1}$ and $\varepsilon_{\text{eff},2}$ are complex effective permittivity of the nanoelements. Considering $\varepsilon_{\text{eff},i} = \varepsilon'_{\text{eff},i} + j\varepsilon''_{\text{eff},i}$ and that $\varepsilon''_{\text{eff},2} \sim 2\varepsilon''_{\text{eff},1}$, while $\varepsilon'_{\text{eff},2} \sim \varepsilon'_{\text{eff},1}$, the reactance of the two impedances is proportional to $X_1 \propto \varepsilon''_{\text{eff},1}/|\varepsilon'_{\text{eff},1} + \varepsilon''_{\text{eff},1}|^2$ and $X_2 \propto 2\varepsilon''_{\text{eff},1}/|\varepsilon'_{\text{eff},1} + 2\varepsilon''_{\text{eff},1}|^2$. Assuming $\varepsilon''_{\text{eff},1} \ll \varepsilon'_{\text{eff},1}$, like in the proposed case which is characterized by low losses ($\tilde{\varepsilon}'' = 0.012j$) and larger real part of the permittivity ($\tilde{\varepsilon}' = 3.6$), the circuit can approximate the behavior of the purely resistive circuit.